\documentclass[prc,showpacs,amsmath,preprint,10pt]{revtex4}
\usepackage{amsfonts}
\usepackage{amsmath}
\usepackage{bm}
\usepackage{epsfig}
\usepackage{mathrsfs}

\begin{document}

 \title {\bf Collective nuclear excitations with Skyrme-Second RPA}

%
%
%

\author{D. Gambacurta,$^{1,2}$\footnote{E-mail: Danilo.Gambacurta@ct.infn.it}
M. Grasso $^{3}$ and F. Catara$^{1,2}$}

\affiliation{$^1$ Dipartimento di Fisica e Astronomia dell'Universit\'a di Catania, Via S.Sofia 64, I-95123 Catania, Italy}
\affiliation{$^2$ Istituto Nazionale di Fisica Nucleare, Sezione di Catania, Via S.Sofia 64, I-95123 Catania, Italy}
\affiliation{$^3$Institut de Physique Nucl\'eaire,
  Universit\'e Paris-Sud, IN2P3-CNRS, F-91406 Orsay Cedex, France}


\begin{abstract}
Second RPA calculations with a Skyrme force are performed to 
describe both high- and low-lying excited states in $^{16}$O. The coupling between 1 particle-1 hole and 2 particle-2 hole as well as that 
between 2 particle-2 hole configurations among themselves are fully taken into account and 
the residual interaction is never neglected,
not resorting therefore to a generally used approximate scheme 
where only the first kind of coupling is considered. The issue of the 
rearrangement terms in
the matrix elements beyond standard RPA is  addressed and discussed.
As a general feature of second RPA results, a several-MeV shift of the 
strength distribution to lower energies is systematically found with 
respect to RPA distributions. A much more important fragmentation of the 
strength is also naturally provided by second RPA due to the huge number of 
2 particle-2 hole configurations.

A better description of the excitation 
energies of the low-lying 0$^+$ and 2$^+$ states is obtained with second RPA
  with respect to 
RPA.

\end{abstract}

\vskip 0.5cm  \maketitle

\section{Introduction}\label{sec:1}
Random Phase Approximation (RPA) is currently used to describe the excitation spectrum of a 
quantum many-body system. Its success in nuclear physics is well 
established and the method has been applied for many years to describe 
giant resonances and low-lying excitation modes. 
However, this approach presents some well known limitations. Extensions and 
procedures to go beyond RPA and to improve the treatment of the correlations present in a 
many-body system  
have been introduced in the 
past decades. A first natural extension is the Quasiparticle RPA (QRPA) where pairing correlations are included by 
defining quasiparticle states through the unitary Bogoliubov transformations \cite{QRPA}. This allows to describe the excitation modes in superfluid open-shell nuclei. 

Other types of correlations may be introduced in a different 
framework. 
A weak point in the formal development of RPA 
 is related to the use of the Quasiboson approximation (QBA) that  
implies a violation of the Pauli principle as well as a severe approximation 
on the reference state: the uncorrelated Hartree-Foch (HF) ground state is used in place of the correlated one. 
An explicitly correlated ground state as reference state is employed 
in those extensions of RPA where either the use of QBA is avoided or its effects are 
cured. Several examples of these beyond RPA methods have been discussed in the past decades 
(see, e.g., \cite{ga09} and references therein). 

Another natural extension of RPA, also based on QBA, is the Second RPA (SRPA) method where 2 particle-2 hole ($2p2h$) excitations 
are included together with the usual RPA 1 particle-1 hole ($1p1h$) configurations providing in this way 
a richer description of the excitation modes. The SRPA equations are well 
known since many years and have been derived by following different procedures. Some examples are the derivation 
within the equations-of-motion method \cite{Yannouleas}, the procedure employing a variational approach 
\cite{Providencia} or the small amplitude limit of the Time Dependent Density Matrix (TDDM) \cite{Tohyama,Lacroix}. 
However, up to very recently, the SRPA equations 
have never been fully and self-consistently solved due to 
the heavy numerical effort they require. Some approximations have been 
adopted in the past, 
namely the SRPA equations have been reduced to a simpler 
second Tamm-Dancoff model (i.e. the matrix $B$ is put equal to zero, see for instance \cite{ho,kn,ni1,ni2}) 
and/or the equations have been solved with uncorrelated 
$2p2h$ states: 
the residual interaction terms in the matrix that couples $2p2h$ 
 configurations among themselves have been 
neglected (diagonal 
approximation) \cite{ad,sc1,sc2,dr1,dr2,yann2,yann3}. 
Very recently, this problem has regained a new 
interest; it is becoming now numerically more accessible and  
the SRPA equations have been solved for closed-shell nuclei using 
an interaction derived from the Argonne V18 potential (with the Unitary 
Correlation Operator Method) 
\cite{roth1,roth2} and for small metallic clusters in the jellium approximation 
\cite{gamba}. 

The increasing interest in the context of 
nuclear structure is also justified
by the manifestation of new phenomena  in unstable nuclei. For instance, 
pygmy resonances represent exotic low-lying excitation modes related to 
the presence of a skin in neutron-rich nuclei. The necessity to go beyond 
standard mean field to describe these resonances has been 
demonstrated by the important effects found when 
particle-vibration coupling is included \cite{ter}. More in general, 
several low-lying excited states may be affected by the particle-phonon 
coupling and in SRPA particle-vibration coupling is 
fully included without any approximation. 
Furthermore, some low-energy excitations like 
the first 0$^+$ or 2$^+$ states in magic nuclei are not reproduced 
by the standard RPA model 
because 
 configurations beyond $1p1h$ are needed to describe them.
Finally, SRPA would allow to study in a proper way the double 
phonon excitation modes that are experimentally well known in nuclei 
\cite{dopho} and have been the object of several theoretical analysis: 
most of these
 studies have been so far based on boson mapping procedures which, however, 
require high order expansions in order to take into account the corrections
due to the Pauli principle (see Ref. \cite{lanza} and references therein).  
In all these cases, interesting results may be expected by the application 
of 
SRPA including all kinds of coupling between $1p1h$ and 
$2p2h$ elementary excitations.

The most currently employed 
phenomenological interactions in nuclear mean field models 
 are density-dependent forces of Skyrme or Gogny type. 
It is well known that, with density-dependent forces, the residual 
interaction used to evaluate the RPA matrices $A$ and $B$ contains a 
rearrangement term coming from the derivative with respect to the density 
of the mean field hamiltonian (second derivative of the energy density 
functional). When dealing with the SRPA problem with density-dependent 
forces, a first formal aspect to consider is the determination of the 
residual interaction that has to be used in the new matrix elements 
with respect to RPA. 
To our knowledge, this aspect has not yet been clarified in the 
literature. A prescription is introduced in Ref. \cite{ad1} and
 rearrangement 
terms appear in the  matrix elements  beyond RPA. On the 
other hand, in more recent calculations \cite{ad2}, 
the same authors have not actually used that prescription and have not 
included 
those rearrangement terms. For the results shown here, we have explored two 
possibilities: i) we have not  
 included rearrangement terms in beyond RPA matrix elements; ii) we have calculated them 
with the usual RPA prescription. 
In a forthcoming article, we will discuss in more detail 
the formal derivation of the residual interaction in the context of SRPA 
with density dependent interactions.

To our knowledge, current versions of SRPA are based on non-interacting $2p2h$ configurations and only the interaction between $2p2h$ and $1p1h$ has been generally taken into account, in the so-called diagonal approximation.
In this work, we present full (the diagonal approximation is not employed and the matrix $B$ is different from zero) 
Skyrme-SRPA results obtained for the doubly magic nucleus 
$^{16}$O. 
In Section II the formal scheme of SRPA is briefly summarized and the 
use 
of QBA in the context of SRPA is commented. 
Numerical checks on stability and sum rules are presented in Sec. III. 
The choice of the numerical energy cutoff is discussed. This 
is a crucial point because the zero-range character of the Skyrme interaction 
generates an ultraviolet divergence in some 
matrix elements.  
Results are shown in Sec. IV 
and the comparison between RPA and SRPA excitation spectra is done 
for both giant resonances and low-lying states in $^{16}$O. 
In particular, for the giant resonances the transition densities are analyzed and the radial distributions related to the main peaks are 
shown for RPA and SRPA.    
For the low-lying states, a special interest is devoted to the 
first 0$^+$ and 2$^+$ excitation modes. They are mainly 
composed by $2p2h$ configurations and, for this reason, cannot be correctly predicted 
by RPA. 
A comparison of the full SRPA 
results with those obtained in the diagonal approximation is also presented. Conclusions and perspectives are finally 
discussed in Sec. V.

\section{Formal scheme}\label{sec:2}
We briefly recall the main formal aspects of SRPA that may be found in several articles (see, for instance, Ref. \cite{Yannouleas}). 
SRPA is a natural extension of RPA where the 
excitation operators $Q^+_{\nu}$ are a superposition of $1p1h$ and $2p2h$ configurations:

\begin{displaymath}
    Q_{\nu}^{\dagger}=\sum_{ph}(X_{ph}^{\nu}a_{p}^{\dag}a_{h}-Y_{ph}^
{\nu}a_{h}^{\dag}a_{p})
\end{displaymath}
\begin{equation}\label{srpa_op}
+    \sum_{p<p',h<h'}(X_{php'h'}^{\nu}
a_{p}^{\dag}a_{h}a_{p'}^{\dag}a_{h'}-Y_{php'h'}^{\nu}a_{h}^{\dag}a_{p}
a_{h'}^{\dag}a_{p'}).
\end{equation}

The $X'$s and $Y'$s are solutions of the equations,
\begin{equation}\label{eq_srpa}
\left(\begin{array}{cc}
  \mathcal{A} & \mathcal{B} \\
  -\mathcal{B}^{*} & -\mathcal{A}^{*} \\
\end{array}\right)
\left(%
\begin{array}{c}
  \mathcal{X}^{\nu} \\
  \mathcal{Y}^{\nu} \\
\end{array}%
\right)=\omega_{\nu}
\left(%
\begin{array}{c}
  \mathcal{X}^{\nu} \\
  \mathcal{Y}^{\nu} \\
\end{array}%
\right),
\end{equation}
where:
\begin{displaymath}
\mathcal{A}=\left(\begin{array}{cc}
  A_{11} & A_{12} \\
  A_{21} & A_{22} \\
\end{array}\right),
\mathcal{B}=\left(\begin{array}{cc}
  B_{11} & B_{12} \\
  B_{21} & B_{22} \\
\end{array}\right),
\end{displaymath}
\begin{displaymath}
\mathcal{X}^{\nu}=\left(\begin{array}{cc}
  X_{1}^{\nu} \\
   X_{2}^{\nu} \\
\end{array}\right),
~~~~\mathcal{Y}^{\nu}=\left(\begin{array}{cc}
  Y_{1}^{\nu} \\
   Y_{2}^{\nu} \\
\end{array}\right).
\end{displaymath}

The indices $1$ and $2$ are a short-hand notation for the $1p1h$ and $2p2h$ configurations, respectively.
 $A_{11}$ and $B_{11}$ are the usual RPA matrices, $A_{12}$ and $B_{12}$ are the matrices coupling 
$1p1h$ with $2p2h$ configurations and $A_{22}$ and $B_{22}$ are the matrices 
coupling $2p2h$ configurations among themselves. 
If QBA is used and the HF ground state is thus employed to evaluate these matrix elements, it can be shown 
that $B_{12}$, $B_{21}$ and $B_{22}$ are zero. The other matrix elements are equal to:

\begin{widetext}
\begin{eqnarray}\label{a12}
A_{12}=A_{ph,p_1p_2h_1h_2}&=&\big\langle  HF |
\big[a_{h}^{\dag}a_{p},[H,a_{p_1}^{\dag}a_{p_2}^{\dag}a_{h_2}a_{h_1}
     ]\big]|  HF \big\rangle =\chi(h_1,h_2)\bar{V}_{h_1pp_1p_2}\delta_{hh_2}-\chi(p_1,p_2)\bar{V}_{h_1h_1p_1h}\delta_{pp_2}
\end{eqnarray}
\begin{eqnarray}\label{a22}
A_{22}=A_{p_1h_1p_2h_2,p'_1h'_1p'_2h'_2}&=&\big\langle HF |\big[a_{h_1}^{\dag}a_{h_2}^{\dag}a_{p_1}a_{p_2},[H,
a_{p'_2}^{\dag}a_{p'_1}^{\dag}a_{h'_2}a_{h'_1} ]\big]|  HF \big\rangle=\nonumber \\
 &=&(\epsilon_{p_1}+\epsilon_{p_2}-\epsilon_{h_1}-\epsilon_{h_2})\chi(p_1,p_2)\chi(h_1,h_2)\delta_{h_1h'_1}\delta_{p_1p'_1}\delta_{h_2h'_2}\delta_{p_2p'_2}+
\chi(h_1,h_2)\bar{V}_{p_1p_2p'_1p'_2}\delta_{h_1h'_1}\delta_{h_2h'_2}+\nonumber\\&&
\chi(p_1,p_2)\bar{V}_{h_1h_2h'_1h'_2}\delta_{p_1p'_1}\delta_{p_2p'_2}+
\chi(p_1,p_2)\chi(h_1,h_2)\chi(p'_1,p'_2)\chi(h'_1,h'_2)\bar{V}_{p_1h'_1h_1p'_1}\delta_{h_2h'_2}\delta_{p_2p'_2}
\end{eqnarray}
\end{widetext}
where the $\epsilon$'s are the HF single particle energies, $\bar{V}$ is the residual interaction and $\chi(ij)$ is the antisymmetrizer for the indices $i$, $j$.

It can be shown \cite{Providencia,dr1}, that the SRPA problem can be reduced to an energy-dependent, RPA-like, eigenvalue problem  but  where the  $A_{11}$ RPA matrix depends  on the excitation energies
\begin{equation}
 A_{1,1'}(\omega)=A_{1,1'}+\sum_{2,2'}A_{1,2}(\omega +i\eta -A_{2,2'})^{-1}{A}_{2',1'}.
\end{equation}
In order to calculate this energy-dependent part one has to invert the $A_{22}$ matrix defined in Eq. (\ref{a22}), whose dimensions are generally very large, requiring thus a strong numerical effort. However, if  the terms depending on the residual interaction  are neglected, resorting to the so-called diagonal approximation, the inversion is algebraic. This approximation, often used in SRPA calculations, will be analyzed in Sections \ref{sec:4}. 

Expressions (\ref{a12}) and (\ref{a22}) are valid in cases where the interaction is not density dependent. Rearrangement 
terms should be included in the case of density-dependent forces. To obtain the results discussed in Sec IV, 
i) we have calculated the matrix elements (\ref{a12})-(\ref{a22}) with $V=V(\rho_0)$  ($\rho_0$ being the HF density) without any rearrangement terms, ii) we have evaluated them 
by adding the usual RPA rearrangement contributions also in the $A_{12}$ and $A_{22}$ matrices. 

It can be shown that the Energy Weighted Sum Rules (EWSR) are satisfied in SRPA \cite{Yannouleas}. Moreover, the first moment
\begin{equation}\label{m1}
 m_1=\sum_{\nu}\omega_\nu |\langle \nu \mid F\mid 0 \rangle |^2
\end{equation}
for a generic one-body operator is found to be the same in RPA and SRPA \cite{Adachi88}. A numerical check of this feature is provided in next section. 

Some comments about the use of QBA in SRPA can be found in the literature \cite{gamba,qbasrpa,srpalipkin}: it is said that the use of QBA 
in SRPA is a more drastic and severe approximation than in RPA. This can be easily understood within the 
variational derivation of SRPA provided by Providencia \cite{Providencia}. 
The usual way of writing the RPA ground state is 

\begin{equation}
|\Psi \rangle = e^{\hat{S}}|\Phi \rangle ~,
\label{gs}
\end{equation}
where :
\begin{equation}
\hat{S}=\sum_{ph} C_{ph}(t)a^{\dagger}_{p}a_{h}
\end{equation} 
the operator $\hat{S}$ being a superposition of $1p1h$ configurations built on top of the HF ground state $|\Phi \rangle$.
The expression of the SRPA ground state in Ref. \cite{Providencia} is the same as that in Eq.   (\ref{gs}) where now the operator $\hat{S}$ also contains  $2p2h$ terms:
\begin{equation}\hat{S}=\sum_{ph} C_{ph}(t)a^{\dagger}_{p}a_{h}+\frac{1}{2}\sum_{php'h'} \hat{C}_{pp'hh'}(t)a^{\dagger}_{p}a^{\dagger}_{p'}a_{h}a_{h'}
 \end{equation}

\begin{figure}
\begin{flushleft}
\epsfig{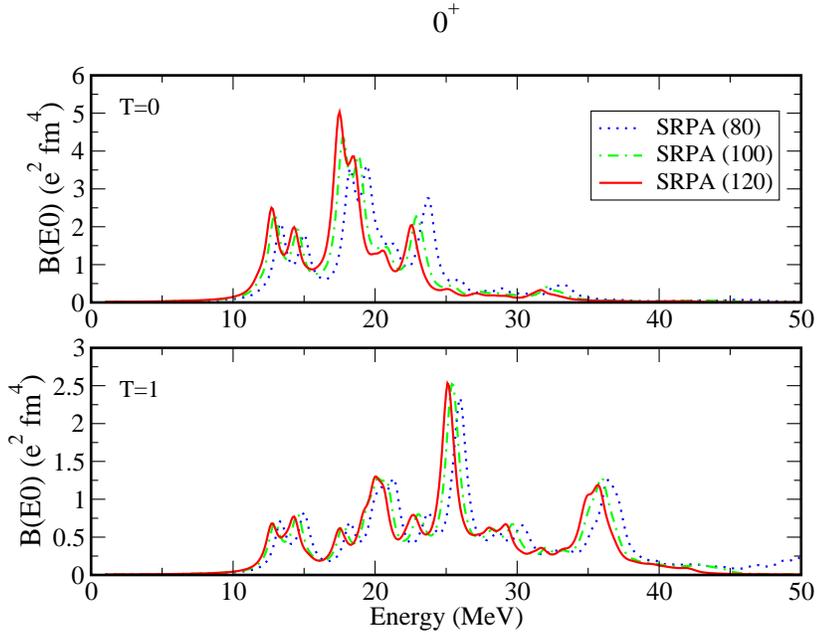} \caption{(Color online) Isoscalar (upper panel) and isovector (lower panel) strength distributions for
  monopole states obtained in SRPA for increasing values of the energy cutoff, indicated in MeV in parenthesis in the figure, on the $2p2h$ configurations.
}\label{evolution-0}
\end{flushleft}
\end{figure}
This means that the ground state is not anymore a Slater determinant. Due to this, the use of the HF ground state to calculate the 
matrix elements (QBA) is a stronger approximation than in RPA.   
Extensions to cure this problem in the context of SRPA have been proposed and applied to a simple model \cite{srpalipkin} and to metal 
clusters \cite{gamba}. Future applications to nuclei are in progress.


\section{Stability of the results and sum rules}\label{sec:3}
In this Section we briefly discuss some technical details of the calculations paying particular attention to the convergence of the results. As a first step, we have solved the HF equation in the coordinate space, by using a 20-fm box. In RPA and SRPA calculations we consider the first $n=7$  single particle (s.p.) states for each  $l$, with $l$ up to $6$. The s.p. wavefunctions have been represented as linear superposition of square well ones.
The SGII \cite{sgii} parametrization of the effective interaction has been used in the present calculations. Since Coulomb and spin-orbit are not taken into account in the residual interaction, our  calculations are   not fully self-consistent  and thus violations of the  EWSR are found (being at worse of $5\%$).
The s.p. space has been chosen large enough to assure that the  EWSR are stable.
In the following we will focus our attention on the excitation spectrum up to 50 MeV.

\begin{figure}
\begin{flushleft}
\epsfig{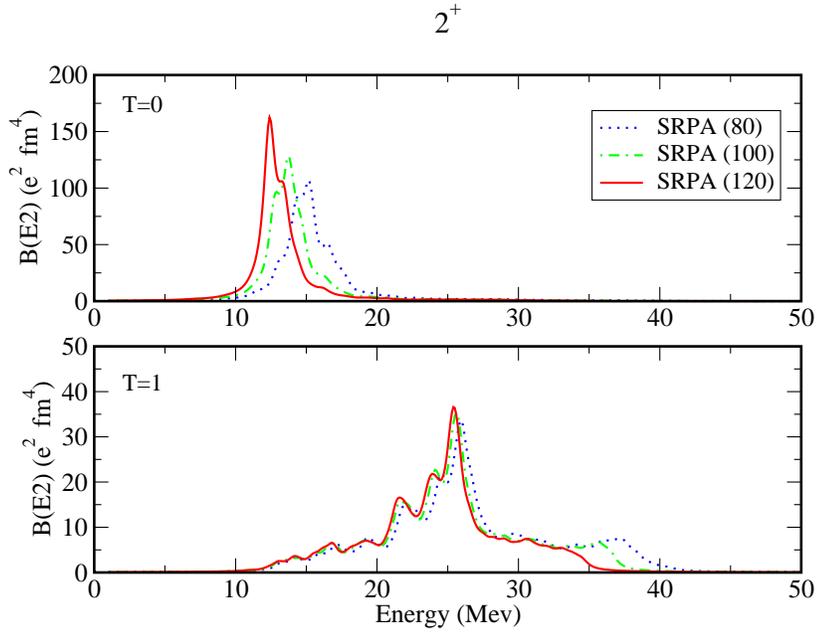} \caption{(Color online) As in Fig. \ref{evolution-0} but for the quadrupole case.}
\label{evolution-2}
\end{flushleft}
\end{figure}

In RPA calculations, $1p1h$ configurations with unperturbed energy up to 100 MeV are considered, while in the SRPA ones, we have considered all the $2p2h$ configurations with an unperturbed energy lower than an energy cutoff  $E_{cut}$.
Due to the ultraviolet divergence related to the zero-range of the 
interaction, results are not expected to converge with respect to the energy 
cutoff. We have however checked that they do not change too drastically in the 
energy regions we are interested in. 
 In order to study the stability of the results with respect to  
the cutoff, we have analyzed how the strength distributions change by 
 increasing  $E_{cut}$  starting from 80 MeV up to 120 MeV.  In 
Figs. \ref{evolution-0} and \ref{evolution-2} 
we show  the monopole and quadrupole strength distributions, respectively,
for different choices of $E_{cut}$ (indicated in 
MeV in parenthesis in the figures). We see that in both cases a 
cutoff equal to 120 Mev is suitable to have stable results. 
Similar stability checks have been systematically 
made for all the results shown in the article.

\begin{figure}
\begin{flushleft}
\epsfig{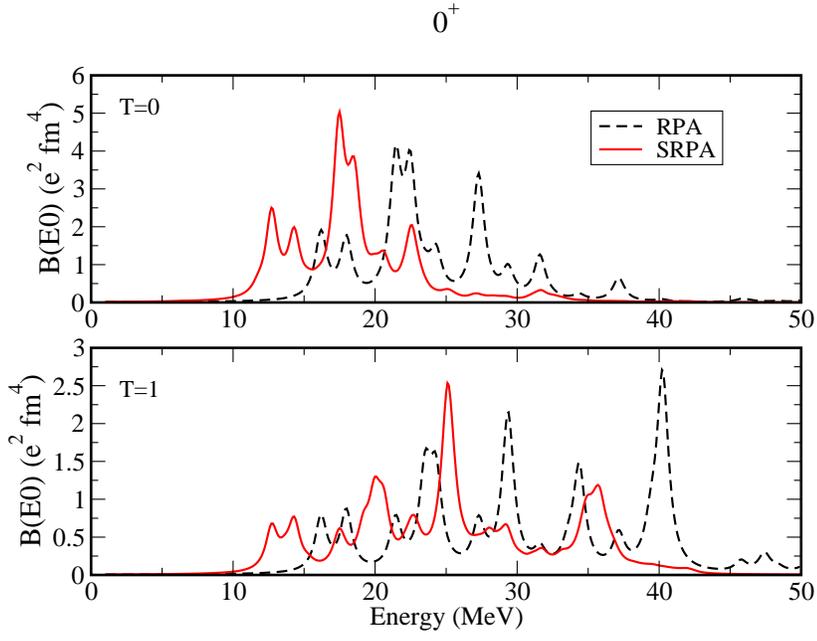} \caption{(Color online) RPA, dashed (black) lines and SRPA, full (red) lines, for the isoscalar (upper panel) and isovector (lower panel)  monopole strength distributions. 
}\label{J0}
\end{flushleft}
\end{figure}

\begin{figure}
\begin{flushleft}
\epsfig{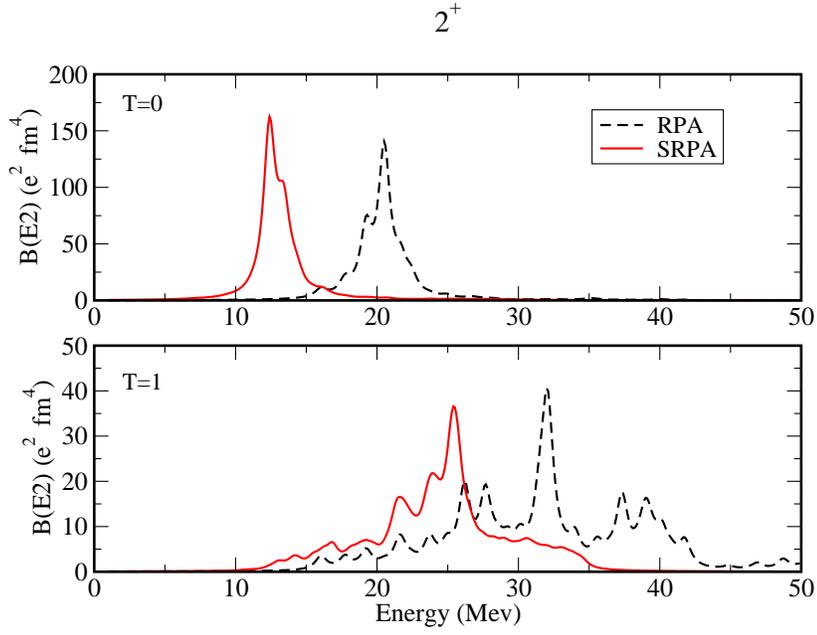} \caption{(Color online) As in Fig. \ref{J0} but for the quadrupole case.}
\label{J2}
\end{flushleft}
\end{figure}

\begin{figure}
\begin{flushleft}
\epsfig{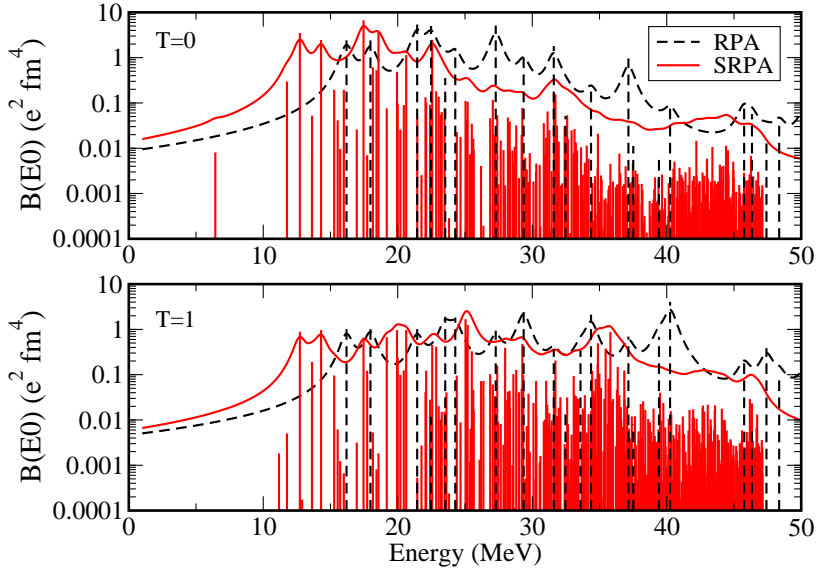} \caption{(Color online) 
As in Fig. \ref{J0} but using a logarithmic scale in the ordinate.}\label{J0-LOG}
\end{flushleft}
\end{figure}
\begin{figure}
\begin{flushleft}
\epsfig{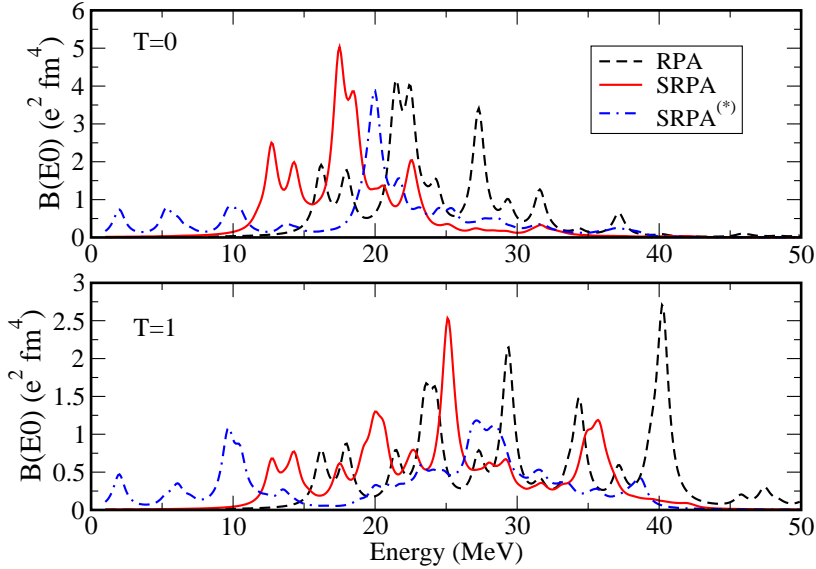} \caption{(Color online) Comparison between RPA, dashed (black) lines, SRPA in full (red) lines and SRPA with rearrangements terms, dot-dashed (blue) lines. The isoscalar (upper panel) and isovector (lower panel)  monopole strength distributions are shown.}. 
\label{J0-NR-R}
\end{flushleft}
\end{figure}

\begin{figure}
\begin{flushleft}
\epsfig{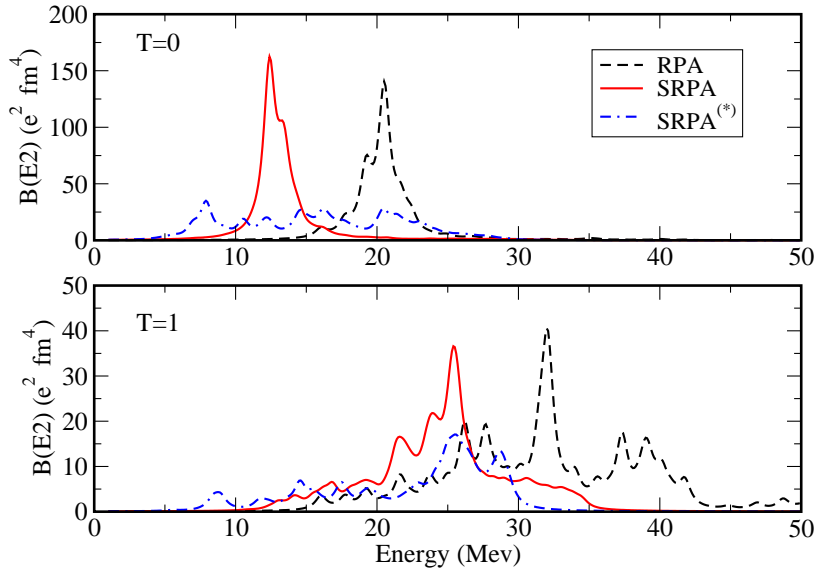} \caption{(Color online) As in Fig. \ref{J0-NR-R} but for the quadrupole case.}
\label{J2-NR-R}
\end{flushleft}
\end{figure}

\begin{figure}
\begin{flushleft}
\epsfig{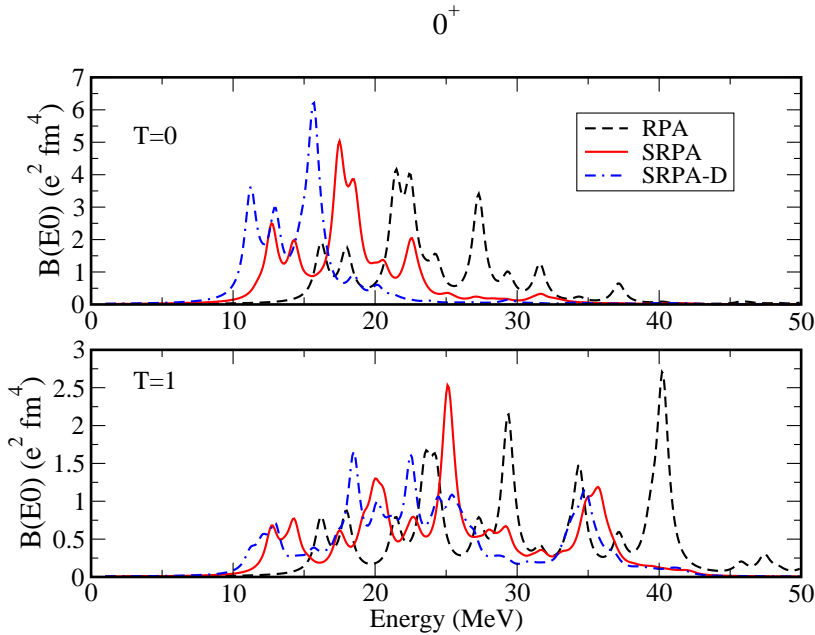} \caption{(Color online) Comparison between RPA, dashed (black) lines, SRPA in full (red) lines and SRPA in the diagonal approximation, dot-dashed (blue) lines. The isoscalar (upper panel) and isovector (lower panel)  monopole strength distributions are shown.}. 
\label{J0-NR-R}
\end{flushleft}
\end{figure}

\begin{figure}
\begin{flushleft}
\epsfig{file=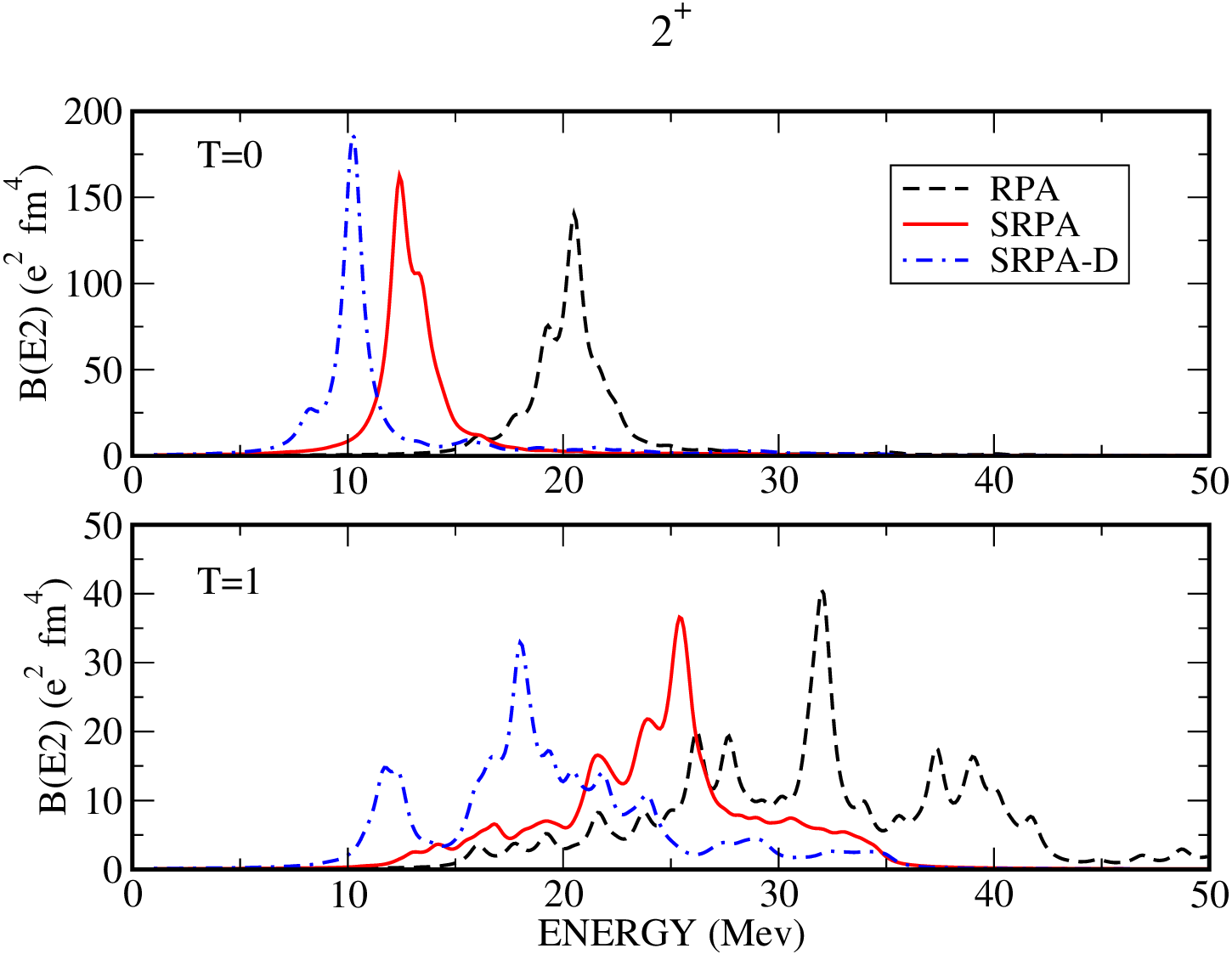,width=0.60\columnwidth,angle=0} \caption{(Color online) As in Fig. \ref{J0-NR-R} but for the quadrupole case.}
\label{J2-NR-R}
\end{flushleft}
\end{figure}

\begin{figure}
\begin{flushleft}
\epsfig{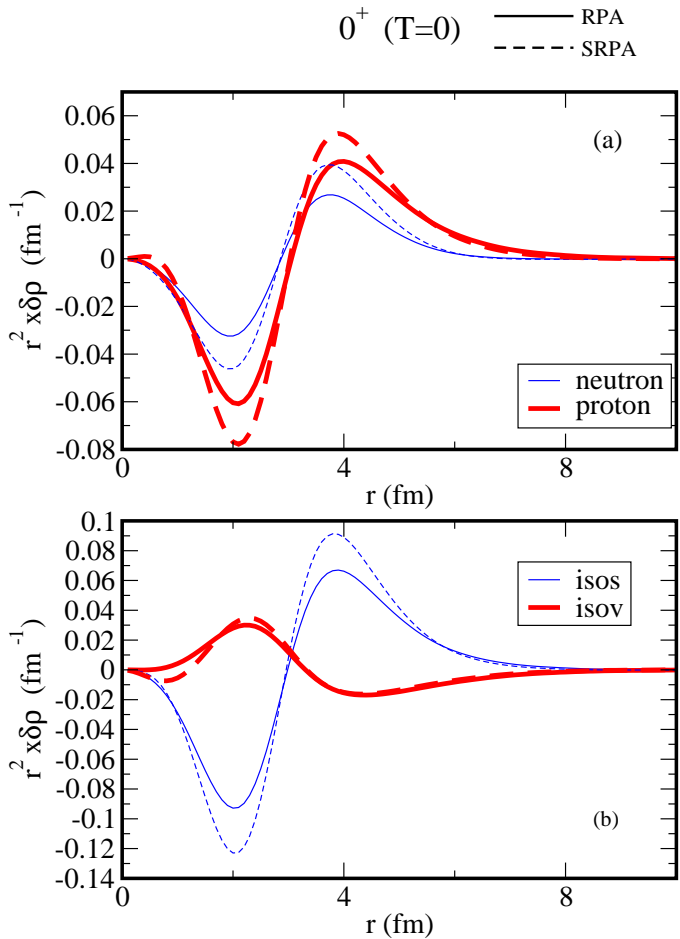} \caption{(Color online) Comparison between RPA (full lines) and SRPA  (dotted lines) transition densities for the monopole isoscalar states (see text): in the panel  (a) neutron
(thin lines) and proton (bold lines) parts, while in panel (b) the isoscalar
(thin lines) and isovector (bold lines) ones. See text for more details.}
\label{TRD-J0T0}
\end{flushleft}
\end{figure}
\begin{figure}
\begin{flushleft}
\epsfig{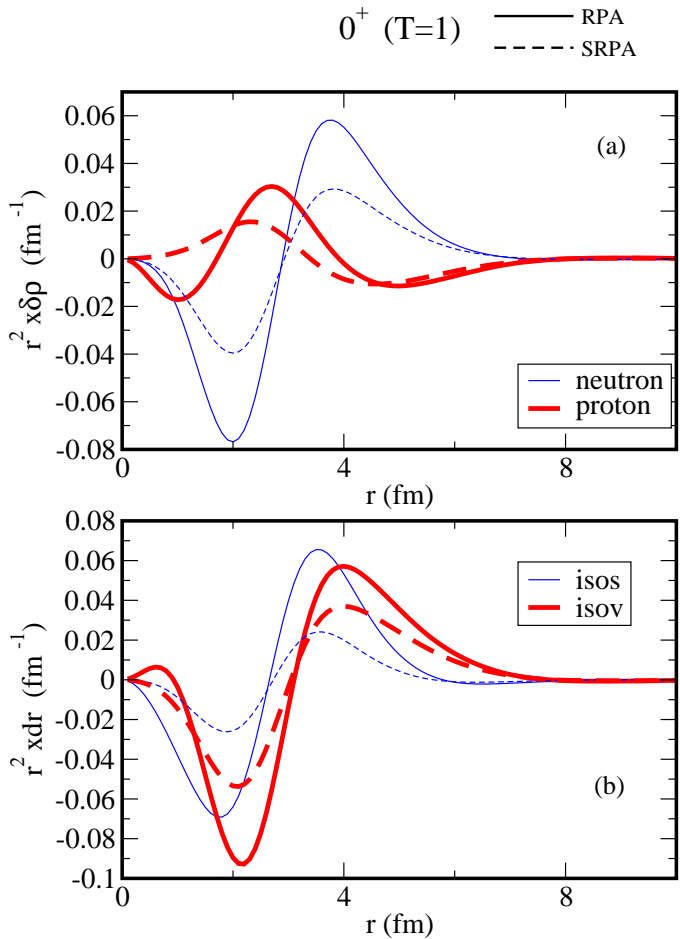} \caption{(Color online) As in Fig. \ref{TRD-J0T0} but for the monopole isovector states.}
\label{TRD-J0T1}
\end{flushleft}
\end{figure}
\begin{figure}
\begin{flushleft}
\epsfig{file=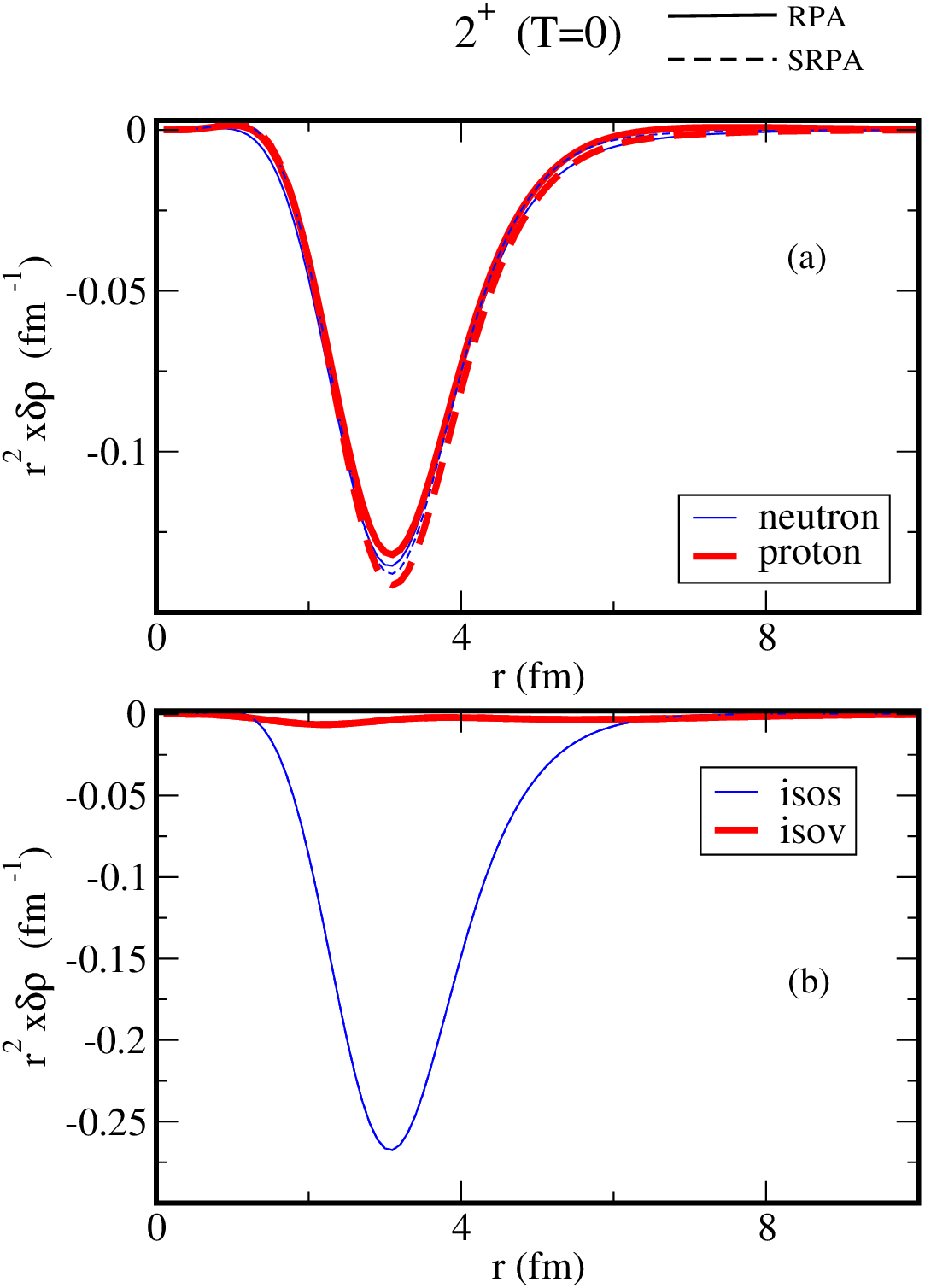,width=0.50\columnwidth,angle=0} \caption{(Color online) Comparison between RPA (full lines) and SRPA  (dotted lines) transition densities for the quadrupole isoscalar states (see text): in the panel  (a) neutron
(thin lines) and proton (bold lines) parts, while in panel (b) the isoscalar
(thin lines) and isovector (bold lines) ones. See text for more details.}
\label{TRD-J2T0}
\end{flushleft}
\end{figure}
\begin{figure}
\begin{flushleft}
\epsfig{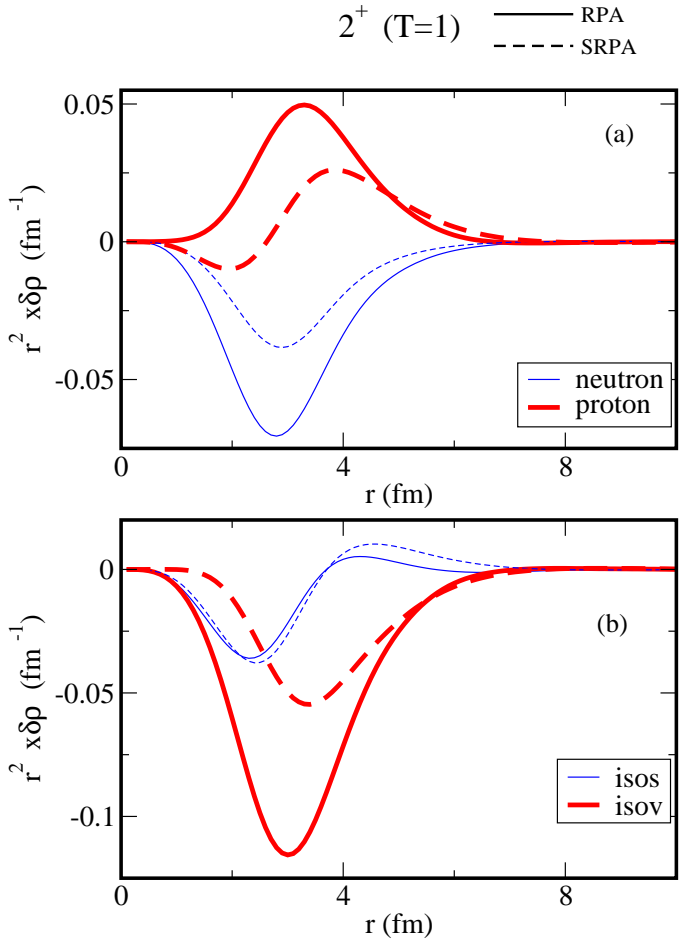} \caption{(Color online) As in Fig. \ref{TRD-J2T0} but for the quadrupole isovector case.}
\label{TRD-J2T1}
\end{flushleft}
\end{figure}

\begin{figure}
\begin{flushleft}
\epsfig{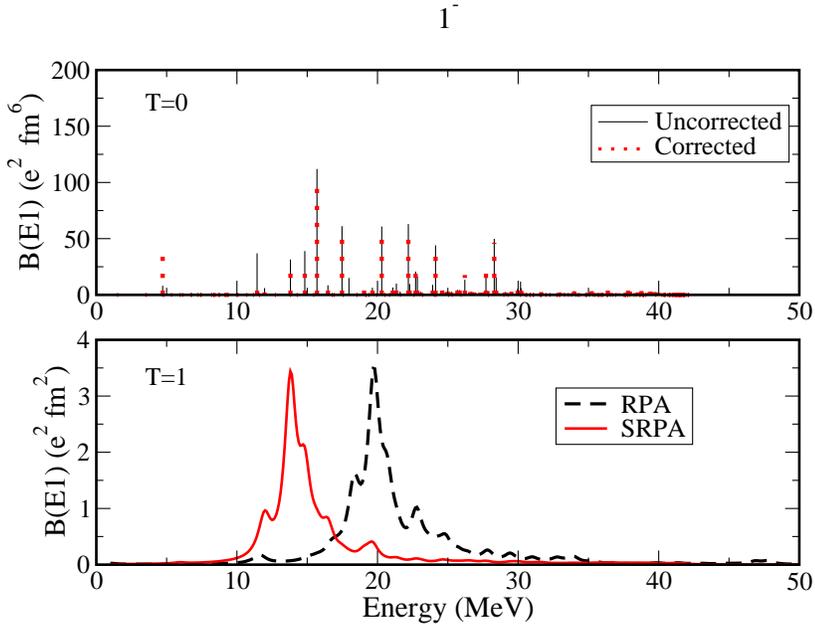} \caption{(Color online) Upper panel: SRPA isoscalar dipole strength distribution using a transition operator of radial form $(\sim r^3)$, full (black) lines and its corrected in dotted (red) lines form $\sim r^3 - \frac{5}{3} \langle r^2 \rangle r$ in order to take into account CM corrections. Lower panel: RPA, dashed (black) line and SRPA in  full (red) line isovector dipole strength distribution using  the standard dipole transition operator of radial form $(\sim r)$. }
\label{J1}
\end{flushleft}
\end{figure}

\begin{figure}
\begin{flushleft}
\epsfig{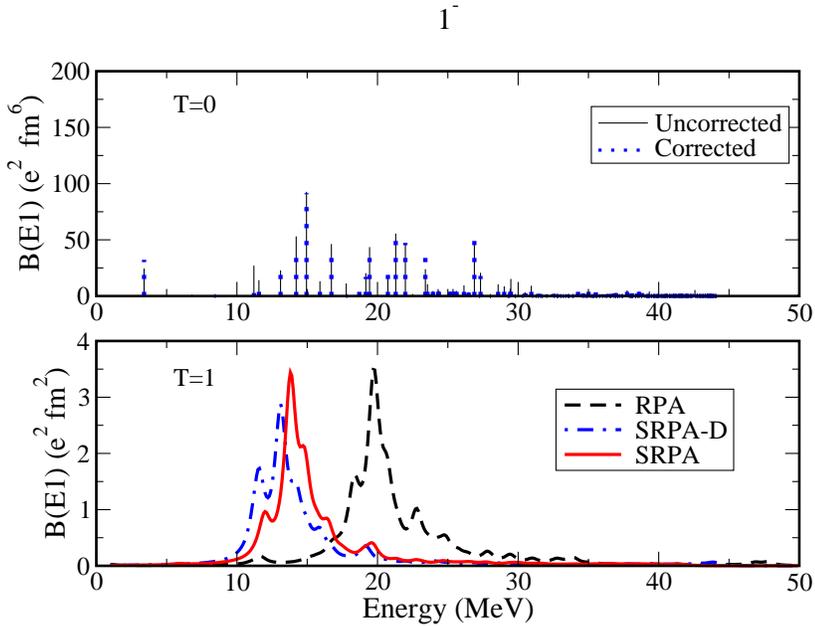} \caption{(Color online) Upper panel:  isoscalar dipole strength distribution obtained within the SRPA in the diagonal approximation using a transition operator of radial form $(\sim r^3)$, full (black) lines and its corrected dotted (blue) lines form $\sim r^3 - \frac{5}{3} \langle r^2 \rangle r$ in order to take into account CM corrections. Lower panel: RPA,  dashed (black) line, SRPA in the diagonal approximation,  dot-dashed (blue) line and  full SRPA in full (red) line isovector dipole strength distribution using  the standard dipole transition operator of radial form $(\sim r)$.   }
\label{J1-DIAG}
\end{flushleft}
\end{figure}

\begin{figure}
\begin{flushleft}
\epsfig{file=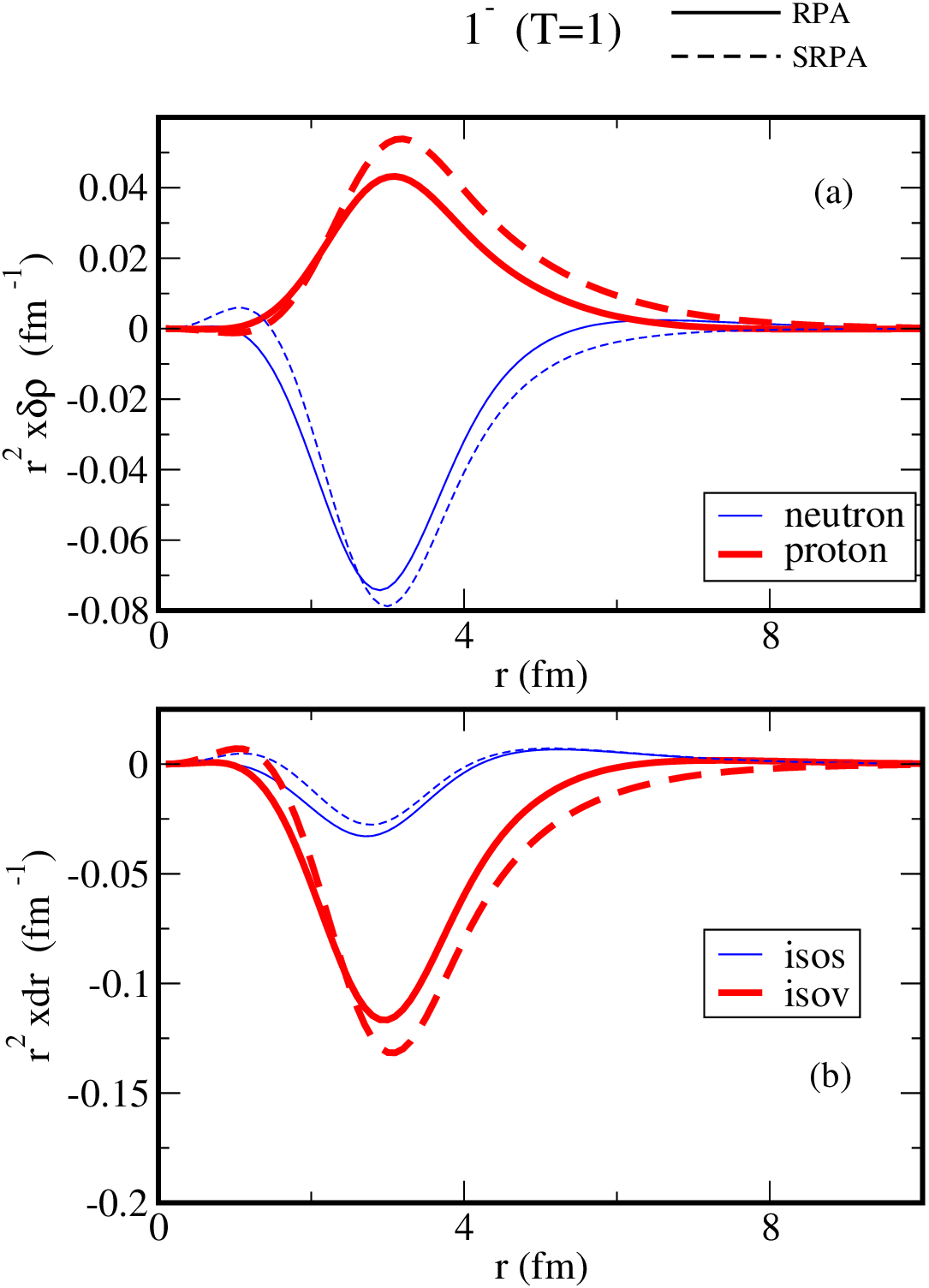,width=0.50\columnwidth,angle=0} \caption{(Color online) As in Fig. \ref{TRD-J2T0} but for the dipole isovector case.}
\label{TRD-J1T1}
\end{flushleft}
\end{figure}

As mentioned above, the EWSR  are satisfied in SRPA  and  the first moment (\ref{m1})
is the same in RPA and SRPA. For a generic one body operator, 
\begin{equation}
 F=\sum_{\alpha,\beta}\langle \alpha\mid F\mid \beta\rangle a_\alpha^\dagger a_\beta, 	\end{equation} 
the transition amplitudes are easily calculated and they have the same expression both in RPA and SRPA, namely
\begin{displaymath}
\langle 0\mid [Q_\nu, F]\mid 0\rangle \approx \langle HF \mid [Q_\nu, F]\mid HF \rangle=
\end{displaymath}
\begin{equation}\label{tra-amp-rpa}
=\sum_{ph}\bigg\{X_{ph}^{\nu*}\langle p\mid F\mid h\rangle+Y_{ph}^{\nu*}\langle h\mid F\mid  p\rangle \bigg\}.
\end{equation}

We note that only the  $p-h$ components of the transition operator  are selected  and that, also in the case of the SRPA, only the  $1p1h$ amplitudes appear in the above equation.

In the present work,  
when a   energy cutoff  $E_{cut}=120 $ Mev on the $2p2h$ configurations 
is used,  SRPA calculations involve the diagonalization 
of large matrices, of the order of $N= 5-6 \times 10^4$ .
A Krylov-Schur iteration procedure from the SLEPC package \cite{slepc} has been used. Since we are interested in the low part of the spectrum, in order to reduce the time of calculation  only the first $n$ eigenvalues (with $n \sim 1-2 \times 10^3$) are calculated. 
In the evaluation of the first moment (\ref{m1}) we  need  to know all the excited states and SRPA calculations with high $2p2h$ configurations  energy cutoff would require a very long calculation time. Therefore,  we have done  some calculations with a smaller energy cutoff, i.e., 
$E_{cut}=60 $ MeV so that  all the energy spectrum can be calculated. In Table \ref{tab:ewsr} we report, for the monopole case, the isoscalar and isovector values , second and third columns respectively, of the SRPA first moment  

\begin{equation}\label{m1srpa}
 m_1=\sum_{\nu}^{\omega_{max}}\omega_\nu |\langle \nu \mid F\mid 0 \rangle |^2
\end{equation}
by including all the states with an excitation energy lower than $\omega_{max}$, whose increasing values are shown in the first column of the table. In the last row, the corresponding RPA values are reported with  $\omega_{max}=100$ MeV. We see that, by increasing the
value of the parameter $\omega_{max}$, the SRPA values of the first moment obtained in SRPA get close to the RPA ones. 
 Similar results have been obtained also in the quadrupole and dipole cases.

We stress that, even for smaller values  of the energy cutoff $E_{cut}$, a similar (almost identical) agreement of the SRPA $m_1$ value with the RPA one is found.
Indeed, by changing  the energy cutoff $E_{cut}$ only a redistribution of the strength is observed while the total strength is the same and is equal to the RPA one. This is related from one hand to the fact that only the  $1p1h$ amplitudes appear in the transition amplitudes and on the other hand to the formal properties of the SRPA equations \cite{Adachi88}.

 \begin{table}\label{Tab:1}
 \begin{center}
\begin{tabular}{|c|c|c|}
\hline
    &$m_1 (T=0)$&$m_1 (T=1)$   \\
\hline
$\omega_{max}$ (MeV) &~~~~~SRPA&SRPA\\
\hline
    40&   626.4381&   115.4153  \\
\hline
    50 &  648.9699&   147.8026 \\
\hline
    60 &  661.0194&   182.7364\\
\hline
    70 &  664.3803&   193.7896\\
\hline
    80 &  669.7185&   197.6874\\
\hline
    90 &  671.4575&   200.6472\\
\hline
   100 &  671.6515&   201.2473\\
\hline
   110 &  671.6515&   201.2473\\
\hline
RPA&671.6516&201.2494\\
\hline

\end{tabular}
\end{center}
\caption{Evolution of the monopole isoscalar and isovector first moments obtained in SRPA, second and third columns respectively, as a function of the $\omega_{max}$ parameter (see Eq. \ref{m1srpa}). In the last row, the corresponding RPA values are reported with  $\omega_{max}=100$ MeV.  }
\label{tab:ewsr}
\end{table}%

\section{SRPA excitation spectrum in $^{16}$O}\label{sec:4}
In this Section we present the nuclear  strength distributions obtained in SRPA for different multipolarities  and we compare them with the RPA ones. The doubly magic nucleus $^{16}$O has been chosen for these first applications of Skyrme-SRPA. 
 At SRPA level different levels of approximation will be considered. As discussed above,  in the case of density-dependent interactions it is not clear how to define the residual interaction appearing in the matrix elements beyond RPA, in particular as far as  the rearrangement terms are concerned. 
In what follows two possibilities have been explored: (i) first, the interaction is used without rearrangement terms in beyond RPA matrix elements;  (i) then, rearrangement terms are included in beyond RPA matrix elements, calculated with the usual RPA prescription. 

Furthermore, the full SRPA calculations are compared with the ones obtained when the diagonal approximation is used.


\subsection{Monopole and  Quadrupole Strength Distributions}
In  this subsection we focus our attention on the monopole and  quadrupole strength distributions. Unless otherwise stated, no rearrangement terms are included in SRPA calculations. In Fig. \ref{J0} we show  the RPA (dashed black lines) and SRPA (full red lines) results for the isoscalar (upper panel) and isovector (lower panel)  monopole strength distributions. In  SRPA all the $2p2h$ configurations with an unperturbed energy lower than an energy cutoff  $E_{cut}=120$ MeV are included.

In both isoscalar and isovector cases the strongest effect in 
SRPA is a several-MeV 
shift of the strength distribution to lower energies with respect to RPA. This 
result seems to be a general feature of SRPA and has been found also 
in different SRPA calculations 
\cite{roth1,roth2,gamba}. A different structure in RPA and SRPA  
strength distribution is also found not leading, however, to 
different widths of the peak profiles in SRPA with respect to RPA.

The same remarks are valid also for the quadrupole case displayed in Fig. 4. 

Fig. 5 gives an idea about how in detail SRPA describes the 
fine structure of the response: a very dense distribution of discrete 
contributions can be seen for the SRPA case due to the existence 
of many 
$2p2h$ elementary excitations in addition to the standard RPA $1p1h$ ones. 

Figs. 6 and 7 represent the same quantities as Figs. 3 and 4, respectively. 
However, this time also the SRPA results obtained with rearrangement terms in 
beyond RPA matrix elements (SRPA$^{(*)}$) are presented in order to evaluate their effect. 

For the isoscalar monopole case (top panel of Fig. 6) the residual interaction 
seems to be more repulsive when rearrangement terms are added providing 
a smaller energy shift to lower energies with respect to RPA. In all the 
other cases shown 
in Figs. 6 and 7, the strength distribution appears very strongly 
fragmented when rearrangement terms are included in all matrix elements. We 
know that the adopted way to evaluate the rearrangement terms in beyond RPA 
matrix elements is not the correct one even if it is currently used. Work is in 
progress to obtain the correct expressions: the fact that the  
SRPA$^{(*)}$ results 
are so different from the SRPA ones underlines that 
this is a very delicate point and indicates that 
the proper expressions are needed. 

In Figs. 8 and 9 the comparison is done with the diagonal approximation. 
The results are quite different and 
this suggests that 
the residual interaction should not be 
neglected in the matrix $A_{22}$ in Skyrme-SRPA 
calculations. 

Finally, Figs. 10 to 13 display the transition densities. 
In Fig. 10 (for the monopole isoscalar response) the SRPA transition 
density refers to the peak at 
$\sim$ 17 MeV while the RPA transition density is evaluated for the peak 
at $\sim$
 21 MeV in top panel of Fig. 3. 
The profiles are not very different meaning that 
the nature of these 
RPA and SRPA excited states is not very different in terms of the spatial 
distributions of wave functions contributing to them. The same considerations
can be done for the results shown in Fig. 12 (referring to the quadrupole 
isoscalar response). In this figure, the RPA (SRPA) 
transition density is calculated for the peak at $\sim$ 22 MeV (13 MeV) 
in top panel of Fig. 4.  

Different is the case for the isovector responses. 
In Fig. 11  the RPA (SRPA) 
transition density, for the monopole response, is calculated for the peak at $\sim$ 29 MeV (25 MeV) shown in the bottom panel of Fig. 3.
For the quadrupole case, the transition density obtained in  the RPA (SRPA) 
 for the peak at $\sim$ 32 MeV (25 MeV), (see lower panel of Fig. 4), are shown in Fig. 13.
  
Important differences are visible between RPA and SRPA suggesting that the 
nature of SRPA and RPA excited states in terms of radial distribution of wave 
functions contributing to the excitation is quite different.

\subsection{Dipole Strength Distributions}
The Thouless theorem on the EWSR \cite{thou} is a very important feature of RPA and it holds also in SRPA \cite{Yannouleas}. It guarantees that spurious excitations corresponding to broken symmetries  separate out and are orthogonal to the physical states. In
 Ref. \cite{SchuckSS} a  detailed discussion about the treatment of single and double spurious modes in RPA and extended RPA theories has been presented. 
In particular, it has been shown that, when 
an approximate ground state is used and/or  $2p2h$ configurations are 
included, all the single-particle amplitudes have to be taken 
into account for the construction of the elementary configurations 
 in order to have single and double spurious modes lying at 
zero energy.
In a self-consistent RPA, i.e., when  the same interaction is used at HF and RPA level, the motion of the center of mass 
(associated with the translational invariance) appears at zero energy being thus exactly separated from the physical spectrum. As mentioned above, our RPA approach is not fully self-consistent and the spurious state lies at about 1 MeV  exhausting more than 96\% of the isoscalar EWSR. In SRPA, as a consequence of the coupling with the $2p2h$ configurations, the spurious state is found to be at imaginary energy. We stress that in a self-consistent SRPA approach this state should appear at zero energy as in RPA. In order to study the possible mixing with spurious components, we have examined the isoscalar dipole strength distribution using a transition operator of the  radial form $(\sim r^3)$ and its corrected form $(\sim r^3 - \frac{5}{3} \langle r^2 \rangle r)$.  
The results are shown in the upper panel of Fig. \ref{J1}. We  can see that some differences between the two cases appear, especially in  the lowest 
part of the spectrum, 
while the mixing with spurious components is smaller in the energy region around and beyond the isovector dipole giant resonance. 
We remind that a prescription often used in the literature for the treatment of the spurious mode consists in multiplying the 
residual interaction by a renormalizing factor such that the spurious mode is found at zero energy. In our RPA calculations  
the spurious state is found at  1.02 MeV and  by using a renormalizing factor of 1.006 its energy goes down to 0.02 MeV; the rest of the isoscalar and isovector distributions remain practically unchanged. The same prescription  
has not been used in SRPA since the use of a renormalizing factor does not solve the problem of the appearance of imaginary solutions, even when larger values of the renormalizing factor are used.

In the lower panel of Fig. \ref{J1} we plot the results for the isovector case, with the excitation operator $(\sim r)$. We observe that going from RPA to SRPA,  as in the monopole and quadrupole cases,   a strong shift towards lower energies of the strength distribution.

In Fig. \ref{J1-DIAG} we show the results when the diagonal approximation is used. In the upper panel of the figure, the isoscalar strength  distributions using the corrected and uncorrected transition operator are shown and we see that, as in the previous case,  very small differences are present when the two different operators are used. In the lower panel of the same figure, the isovector distributions obtained in RPA (dashed-black line), SRPA in the diagonal approximation  (dot-dashed blue line) and  full SRPA (full-red line) are shown.
It is interesting to note that the results obtained in the diagonal approximation are different from the ones obtained in the  full SRPA, the differences however being smaller than the ones found in the monopole and quadrupole cases. 

As far as the rearrangement terms are concerned, we have found 
 in the dipole case
that, even when small values  of the $2p2h$ energy cutoff $E_{cut}$ are used (60-80 MeV), the SRPA equations give imaginary solutions if the rearrangement 
terms are included in all matrix elements. 
Moreover, by comparing in this case 
the isoscalar strength  distributions obtained by using the corrected and uncorrected transition operator, 
very large differences are found, especially in the low energy part of the spectrum, indicating that a strong mixing with spurious components
is present in this case.  In the isovector channel, the position of the main peak is not very different from the one obtained when no rearrangement terms are used (lower panel of Fig. \ref{J1}) but the height of the peak is 
strongly reduced (more than 50\%).

In Fig. \ref{TRD-J1T1}  the RPA (SRPA) 
transition density, for the isovector dipole response, is calculated for the peak at $\sim$ 20 MeV (14 MeV) shown in the bottom panel of Fig. \ref{J1}.
We see that in 
the dipole case the transition densities in RPA and SRPA are quite similar.
This behaviour is at variance  from what we found in the monopole and 
quadrupole cases, as sown in Figs 11 and 13, respectively.


\subsection{Low-lying 0$^+$ and $2^+$ states}

The results for the 0$^+$ and $2^+$ low-lying states obtained in SRPA are
 shown in Tables \ref{Tab:Low0p} and \ref{Tab:Low2p}. These states 
are mainly composed by $2p2h$ configurations. 

The largest $1p1h$ configuration in the $0^{+}$ state is
 $(2p_{\frac{1}{2}},1p_{\frac{1}{2}})^{\pi}$ with an unperturbed energy of 16.17 Mev when no rearrangement terms are considered and $(3p_{\frac{1}{2}},1p_{\frac{1}{2}})^{\nu}$ with an unperturbed energy of 24.03 MeV when rearrangement terms are included. In both cases,
the lowest $2p2h$ configuration is $\big[ [1d_{\frac{5}{2}},1p_{\frac{1}{2}} ]^3[(1d_{\frac{5}{2}},1p_{\frac{1}{2}}]^3 \big]_{\pi}^{0}$ with an unperturbed energy of E=15.26 MeV. 

For the $2^{+}$ state the most important $1p1h$ configuration is $(2d_{\frac{5}{2}},1s_{\frac{1}{2}})^{\pi}$   with an unperturbed energy of 28.28 MeV
when rearrangement terms are not considered. 
When rearrangement terms are included the most important 
configuration is $(2p_{\frac{3}{2}},1p_{\frac{1}{2}})^{\nu}$ with an unperturbed energy of  21.71 MeV.
The lowest $2p2h$ configuration is, in both cases,
$\big[ [1d_{\frac{5}{2}},1p_{\frac{1}{2}} ]^3[(1d_{\frac{5}{2}},1p_{\frac{1}{2}}]^3  
\big]_{\pi}^{2}$ with an unperturbed energy of 15.26 MeV.

       \begin{table}
\begin{center}
\begin{tabular}{|c|c|c|c|c|c|c|}
\hline
\multicolumn{6}{|c|}{Low Lying $0^{+}$ energy (MeV)} \\
\hline
Exp &RPA& SRPA&SRPA-D&SRPA$^*$&SRPA$^*$-D   \\
\hline 
$\sim$6& 16.19&6.43&11.23& 5.29&Imm.	\\
\hline
\end{tabular}
\end{center}
\caption{Energy of the lowest $0^{+}$ state obtained in RPA and in SRPA, compared with the experimental value \cite{Low0p2p}. The results identified with the star are obtained by taking into account rearrangement terms (last two rows).
 With ``SRPA-D'' we indicate the SRPA results when the diagonal approximation is used (fourth and sixth rows). The SRPA result in the diagonal approximation whit rearrangement terms gives imaginary solution (last row). }\label{Tab:Low0p}

\end{table}

\begin{table}
\begin{center}
\begin{tabular}{|c|c|c|c|c|c|c|}
\hline
\multicolumn{6}{|c|}{Low Lying $2^{+}$ energy (MeV) } \\
\hline
Exp &RPA& SRPA&SRPA-D&SRPA$^*$&SRPA$^*$-D   \\
\hline 
$\sim$7& 16.03&7.16&12.44&4.70&Imm.	\\
\hline
\end{tabular}
\end{center}
\caption{As in Table \ref{Tab:Low0p} but for $2^{+}$ state.}\label{Tab:Low2p}
\end{table}

Several conclusions may be drawn about these results. 
First, RPA is not at all able to describe these low-lying states simply because
 beyond $1p1h$ configurations are necessary to construct them. RPA energies 
are indeed far too high in both cases. 

It is striking that the  
SRPA energies are very close to the experimental results. The residual 
interaction seems to be very important for describing these states: the first 
unperturbed $2p2h$ configuration is actually located at about 15 MeV and the 
residual interaction is thus responsible for the strong shift to lower 
energies in the response. When rearrangement terms are included, the shift 
is even stronger. 

The diagonal approximation looks very poor for the treatment of these 
low-energy states indicating that the interaction between $2p2h$ 
configurations is very important for providing the correct excitation energies.

Finally, it can be interesting to compare the SRPA results with other types of 
analysis. As an illustration, let us consider the energy of the first 
0$^+$ excited state. 
Ab-initio coupled cluster investigations do not reproduce at all the energy 
of this state providing an excitation energy of 19.8 MeV \cite{wl}. 
Brown and Green have described the low-lying states in $^{16}$O 
by mixing spherical and deformed states \cite{br}. 
Shell model can nicely describe these states due to the configuration mixing 
\cite{zu}.
Cluster models are also able to well describe this state by assuming an 
$\alpha$+$^{12}$C or a 4$\alpha$ structure  
for the nucleus $^{16}$O (see Ref. \cite{fu} and references therein). 
SRPA is the only RPA-like approach in spherical symmetry 
that 
reproduces this energy without any special modelization for the structure 
of the nucleus $^{16}$O. 

It is worth mentioning however that a complete analysis of these low energy 
states would need also the evaluation of the B(E$\lambda$) values. Work is 
in progress in this direction to evaluate  
the transition probabilities in these cases where the excitations are mainly 
composed by $2p2h$ configurations.

\section{Conclusions}\label{sec:5}

We have performed Skyrme-SRPA calculations for describing collective and 
low-lying excited states in $^{16}$O. The Skyrme interaction SGII is used. 
The SRPA scheme is fully treated without employing the currently used 
Second-Tamm-Dancoff or diagonal approximations. The rearrangement terms of 
the residual interaction are treated in this work (i) by neglecting them 
in the matrix elements beyond RPA; (ii) by calculating them with the usual 
RPA procedure for all the matrix elements. Work is in progress to derive 
the proper expressions to be used in beyond RPA matrix elements. 

A general feature of the SRPA spectra is a several-MeV shift to lower energies 
with respect to RPA distributions. 
This shift being very strong, Skyrme-SRPA energies of giant resonances 
are typically too low with respect to the experimental response 
(Skyrme-RPA results are in general in good agreement with the experimental 
data for these excitations). 
To cure this 
problem we plan to explore the possibility of using an extended SRPA scheme
in the same line of Ref. \cite{gamba,gamba2010}. A longer term project is to look for some 
new parametrization of the effective interaction, adjusted so to make it
suitable for this kind of calculations

The SRPA energies of the low-lying 0$^+$ and 2$^+$ states are in very good 
agreement with the experimental results. Work is in progress to evaluate the 
transition probabilities and complete the analysis of these excitation 
modes. 

Finally, after this first numerical application of the method, we plan 
in future works to  
explore with SRPA heavier or more exotic nuclei and to check  
the model dependence of the results by employing 
also other Skyrme parametrizations.

\begin{acknowledgments}
The authors gratefully thank N. Pillet, P.Schuck, N. Van Giai and M. Tohyama for fruitful  discussions.
\end{acknowledgments}

\end{document}